\documentclass[draftclsnofoot,12pt,onecolumn]{IEEEtran}

\usepackage {graphicx}
\usepackage {amsmath}
\usepackage {amssymb}
\usepackage {ulem}
\usepackage {latexsym}
\usepackage {multirow}
\usepackage{array}
\usepackage[dvips]{epsfig}
\usepackage{subfigure}
\usepackage {enumerate}

\begin{document}

\title{Self-optimized Coverage Coordination in Femtocell Networks}

\author{Han-Shin Jo,~
        Cheol Mun,~
        June Moon,~
        and Jong-Gwan Yook
\thanks{Han-Shin Jo and Jong-Gwan Yook are with Department of Electrical \& Electronic Engineering, Yonsei University,
        Seoul, Korea 120-149. (e-mail: \{gminor, jgyook\}@yonsei.ac.kr).
Cheol Mun is with the Dept. of Electronic Communication
Eng., Chungju National University, Chungju, Korea.
(e-mail:chmun@cjnu.ac.kr).June Moon is with Telecommunication R\&D Center, Samsung Electronics, Suwon, Gyeonggi, Korea
442-742. (e-mail:june.moon@samsung.com).}}

\markboth{} {Shell \MakeLowercase{\textit{et al.}}: Bare Demo of
IEEEtran.cls for Journals} \maketitle
\begin{abstract}
This paper proposes a self-optimized coverage coordination scheme
for two-tier femtocell networks, in which a femtocell base station
adjusts the transmit power based on the statistics of the signal and
the interference power that is measured at a femtocell downlink.
Furthermore, an analytic expression is derived for the coverage
leakage probability that a femtocell coverage area leaks into an
outdoor macrocell. The coverage analysis is verified by simulation,
which shows that the proposed scheme provides sufficient indoor
femtocell coverage and that the femtocell coverage does not leak
into an outdoor macrocell.
\end{abstract}

\IEEEpeerreviewmaketitle

\section{Introduction}
Femtocell technology has been emerging as a solution to the increase
of both capacity and coverage while reducing both the capital
expenditures and operating expenses of cellular networks.
As femtocells share spectrum with macrocell networks, controlling
the cross-tier interference between femto- and macrocells is need to
be considered first in the enhancement of coverage and capacity. In
addition, since a network operator may not be able to control
femtocell locations, it is necessary for femtocells to sense the
radio environment around them and carry out the self-configuration
and self-optimization \cite{SOC1}-\cite{SOC3} of radio parameters
from the moment they are set up by a consumer.

Conventional dynamic cell sizing schemes, which adjust the transmit
power of a base station (BS) \cite{DynamicCell1}\cite{DynamicCell3}
or both the transmit power and antenna beam forming
\cite{DynamicCell2}, have been developed to improve the overall
system capacity compared to that of a fixed cell sizing scheme. 
These approaches are not suitable for a macro/femto overlaid cell
structure, where femtocell coverage must be controlled so it does
not interfere with the outdoor macrocell. In order to achieve this
goal, Claussen \textit{et al}. \cite{DynamicCell4} proposed a
femtocell coverage coordination method that adjusts the femtocell
pilot power, based on the number of handover events from outdoor
passing, and the indoor users, which is robust against the varying
size and shape of buildings. However, outdoor users may already
experience inferior link quality during the time that a femtocell BS
reduces its transmit power after recognizing the handover events of
outdoor users. In particular, in a private access scenario that
serves only registered users, the unauthorized users near a
femtocell have a serious increase in the call drop rate or reduction
of data rate. Moreover, the procedure that reconnects the rejected
outdoor user to the macrocell may induce an additional handover,
which causes a considerable amount of data transmission delay as
well as packet loss in a packet switched cellular network with hard
handovers, such as IEEE 802.16e WiMAX and HSDPA
\cite{HOwimax}\cite{HOhsdpa}. Therefore, this problem is severe for
delay- and packet-loss-sensitive real-time applications such as
Voice over Internet Protocol (VoIP).

This paper proposes a coverage coordination scheme that is based on
the statistics of the signal and interference power measured at a
femtocell downlink (as opposed to the scheme based on handover
events) to prevent any handover of outdoor users in advance. The
proposed scheme comprises both the self-configuration and
self-optimization of femtocell pilot transmit power. With a
self-configuration function, a femtocell BS initiates its transmit
power based on the measurement of interference from neighboring BSs
in a manner that achieves a roughly constant cell coverage. The
femtocell BS then performs a self-optimization function that
continually adjusts the transmit power so that the femtocell
coverage does not leak into an outdoor area while sufficiently
covering an indoor femtocell area.

\section{Downlink Transmit Power Control}\label{sec:TPC}

This study considers a two-tier cellular network composed of
overlaid macrocells and underlaid femtocells in which both cells use
the same frequency channel. A femtocell BS is located at the center
of a building with radius $r_b$. Both BS and user are equipped with
an omni-directional antenna. The femtocell BS creates cell coverage
with radius $r_f$, which is adaptively adjusted by the proposed
transmit power control in order to correspond to the building area
i.e., $r_f=r_b$. The transmit power control is composed from a two
step procedure, where the femtocell BS initially self-configures its
power and self-optimizes cell coverage by using transmit power
control based on the measurements of radio environments.

\subsection{Initial Self-configuration}
A femtocell BS measures the average received power of pilot (over
multiple frames to average out fast fading effect) from the
neighboring macrocell and femtocell BSs on a neighboring BS list.
The femtocell then chooses the strongest pilot power, $I_{b,max}$,
among them. The femtocell BS configures its transmit pilot power
such that the received pilot power from the femtocell BS and the
strongest macrocell BS are identical on average at an initial cell
radius of $r_{ini}$, i.e., $r_f=r_{ini}$. The Appendix shows that
$I_{b,max}$ is nearly identical to the interference power from the
strongest macrocell both measured by and averaged over the users
located at the initial cell edge. Thus, the initial femtocell pilot
power $P_{f,ini}$ (dBm) is determined such that the femtocell BS
power received at $r_{ini}$ is equal to $I_{b,max}$, as follows:
\begin{equation}
P_{f,ini}=\min\left(I_{b,max}+L\left(r_{ini}\right),
P_{\max}\right).\label{eq:InitPw}
\end{equation}
Here, $P_{\max}$ and $L$ are maximum femtocell pilot power and path
loss, respectively. The initial self-configuration only provides the
initial cell coverage of a femtocell, which is refined by the
following self-optimized power control.


\subsection{Self-optimized Power Control}
A femtocell BS measures the level of other-cell interference
$I_u(0)$ that is received from neighboring macrocells and femtocell
BSs. The femtocell BS evaluates the received interference-plus-noise
power $Z_u(0)=10\log_{10}(I_u(0)+W)$, where $W$ is the thermal noise
power. The femtocell BS collects the time-averaged received power
$Q^{(i)}$ in dBm, which is measured by each femtocell user during
the $i$th iteration and is fed back to the currently linked
femtocell BS.
Based on the decision variable\footnote{$\Gamma^{(i)}$ 
gives a rough measure of spatially averaged carrier to
interference-plus-noise ratio (CINR) over a femtocell area.}
$\Gamma^{(i)}=\overline{Q}^{(i)}-Z_u(0)$, where $\overline{Q}^{(i)}$
is the averaged $Q^{(i)}$ over femtocell users, the transmit pilot
power of the femtocell BS at the $i$th iteration is updated by
\begin{equation}
P_f^{(i+1)}=\left\{
                     \begin{array}{ll}
                       \min\left(P_f^{(i)}+\Delta P,P_{\max}\right) & \textrm{for~~}  \Gamma^{(i)}\leq\Gamma_{th}, \\
                       \min\left(P_f^{(i)}-\Delta P,P_{\max}\right) & \textrm{otherwise}.
                     \end{array}
                     \right.
                     \label{eq:Update1}
\end{equation}
Here, $\Delta P$ is the power control step in dB. $P_f^{(i+1)}$ is
determined by comparing $\Gamma^{(i)}$ with a threshold
$\Gamma_{th}=\Gamma_{0}+\Gamma_{\Delta}$, where $0\leq
\Gamma_{\Delta} \leq\Gamma_{\Delta,\max}$. In order to make this
power control scheme work properly, it is essential to set the
threshold appropriately, that is, determining $\Gamma_{0}$ and
$\Gamma_{\Delta,\max}$.
\subsubsection{Statistical Threshold $\Gamma_{0}$}
$\Gamma_{0}$ is obtained from the statistical characteristics of
$\Gamma^{(i)}$. Under the approximation that femtocell users are
uniformly distributed over a circular femtocell with a radius of
$r_f$ at the $i$th iteration, the probability density function (PDF)
of random variable $D$, which represents the distance between the
femtocell BS and user, is given as
$f_D(d)=2d\big{/}({r_f^{(i)}}^2-\varepsilon_0^2),
~d\in[\varepsilon_0,{r_f^{(i)}}^2]$,
where $\varepsilon_0$ is the minimum $D$. Here, all radii and
distances are in meters. Both the outdoor and indoor path loss in dB
are modeled as $L(D)=A_s+10n\log\left(D/d_s\right)$, where $n$ and
$A_s$ denote the path loss exponent and the path loss at a reference
distance of $d_s=1$ m, respectively. For the outdoor-to-indoor path
loss, the wall penetration loss $L_p$ is added to $L(D)$. When
$r_f^{(i)}\leq r_b$, the time-averaged received power of a femtocell
user is given by $ Q^{(i)}=P_f^{(i)}-L\left(D\right)$. Then, from
the PDF of $D$, the PDF of $Q^{(i)}$ is given as
\begin{equation}
f_{Q^{(i)}}(q)=10^{\frac{P_f^{(i)}-A_s-q}{5n}}\ln10\Big/(5n({r_f^{(i)}}^2-\varepsilon_0^2)),
~q\in\left[P_f^{(i)}-L(r_f^{(i)}),
P_f^{(i)}-L(\varepsilon_0)\right].\label{eq:PDF_R}
\end{equation}
The expected value of $Q^{(i)}$ is given by
\begin{equation}
E\left[Q^{(i)}\right]=5n/\ln10+P_f^{(i)}-A_s-10n({r_f^{(i)}}^2\log
r_f^{(i)}-\varepsilon_0^2\log\varepsilon_0)
\Big/({r_f^{(i)}}^2-\varepsilon_0^2). \label{eq:E_R}
\end{equation}
Let denote $\overline{I}_u(D)$ as the other-cell interference averaged over the users uniformly located on a circumference of radius of $D$ centered at their femtocell BS. Since $I_u(0) \approx \overline{I}_u(D)$ for $D < 0.5 d_{b,i}$ from the Appendix, $I_u(0)\approx \overline{I}_u(r_f^{(i)})$; therefore,
$\Gamma^{(i)}$ is approximated as
\begin{equation}
\Gamma^{(i)}=\overline{Q}^{(i)}-Z_u(0)\approx
E\left[Q^{(i)}\right]-\overline{Z}(r_f^{(i)}), \label{eq:Gamma1}
\end{equation}
where
$\overline{Z}(r_f^{(i)})=10\log_{10}\left(\bar{I}_u(r_f^{(i)})+W\right)$.
Furthermore, when $r_f^{(i)}<r_b$, the average CINR of the user
located at a femtocell edge is equal to a CINR threshold
$\gamma_{th}$, i.e.,
$Q^{(i)}(r_f^{(i)})-\overline{Z}(r_f^{(i)})=\gamma_{th}$. From
this constraint, $\Gamma^{(i)}$ can be further approximated as
follows:
\begin{eqnarray}
\Gamma^{(i)}&\approx&E\left[Q^{(i)}\right]-Q^{(i)}(r_f^{(i)})+\gamma_{th}\cr
&\mathop =\limits^{\left( a \right)}&
5n/\ln10+\gamma_{th}-\tfrac{10n\varepsilon_0^2}{{r_f^{(i)}}^2-\varepsilon_0^2}\log\tfrac{r_f^{(i)}}{\varepsilon_0},
~\textrm{for}~~r_f^{(i)}<r_b, \label{eq:Gamma2}
\end{eqnarray}
where (a) follows from (\ref{eq:E_R}) and the equation $ Q^{(i)}=P_f^{(i)}-L\left(D\right)$.
This interestingly shows that $\Gamma^{(i)}$ depends only on
$r_f^{(i)}$, and increases and converges to
$\Gamma_0=\frac{5n}{\ln10}+\gamma_{th}$ as $r_f^{(i)}$ increases
(see the $\Gamma^{(i)}$ graph $(i\leq15)$ in Fig. \ref{fig:Gamma}).

Fig. \ref{fig:Gamma} describes an example of the coverage adaptation
process that uses the power control scheme with
$\Gamma_{th}=\Gamma_0$. When $r_f^{(i)}<r_b~(i<15)$, $\Gamma^{(i)}$
is determined by (\ref{eq:Gamma2}) and it is less than $\Gamma_0$.
Therefore, as the iteration index $i$ increases, the transmit power
$P_f^{(i)}$ increases, and the femtocell coverage $r_f^{(i)}$
extends to a building wall. The first time that $r_f^{(i)}$ is equal
to $r_b~(i=15)$, $\Gamma^{(15)}$ is a little less then $\Gamma_0$,
which leads to an increase in $P_f^{(i)}$. However, contrary to the
case of $i<15$, the increase of transmit power no longer extends the
femtocell coverage to the outdoors, i.e., $r_f^{(i+1)}=r_f^{(i)}$,
until the transmit power becomes large enough to overcome the
additional path loss due to wall penetration so that femtocell
coverage leaks into the outdoor region (see Case 2 in Fig.
\ref{fig:Margin}). This constant cell coverage causes constant average interference-plus-noise power at the cell edge, i.e.,
$\overline{Z}(r_f^{(i+1)})=\overline{Z}(r_f^{(i)})$, thus
$\Gamma^{(i+1)}-\Gamma^{(i)}=E[Q^{(i+1)}]-E[Q^{(i)}]$ from
(\ref{eq:Gamma1}). Additionally, $r_f^{(i+1)}=r_f^{(i)}$ results in
$E[Q^{(i+1)}]-E[Q^{(i)}]=P_f^{(i+1)}-P_f^{(i)}$ from (\ref{eq:E_R}),
and $P_f^{(i+1)}-P_f^{(i)}=\pm\Delta P$ from (\ref{eq:Update1}),
i.e., $\Gamma^{(i+1)}-\Gamma^{(i)}=\pm\Delta P$. Therefore,
$\Gamma^{(i)}$ is reformulated as
\begin{equation}
\Gamma^{(i+1)} =\left\{
                     \begin{array}{ll}
                       \Gamma^{(i)}+\Delta P & \textrm{for}~~\Gamma^{(i)}\leq\Gamma_{th},\\
                       \Gamma^{(i)}-\Delta P &
                       \textrm{otherwise}
                     \end{array}\right. ~\textrm{for}~~r_f^{(i)}=r_b.\label{eq:Gamma3}
\end{equation}
According to this equation, $\Gamma^{(16)}=\Gamma^{(15)}+\Delta
P>\Gamma_0$ on the assumption that $\Delta P>\Gamma_0-\Gamma^{(15)}$
(more iterations will provide $\Gamma^{(i)}>\Gamma_0$ when $\Delta
P<\Gamma_0-\Gamma^{(15)}$), and $P_f^{(i)}$ increases no more than
$P_f^{(16)}$. Thus, $\Gamma_{th}$, which is set to $\Gamma_0$, provides
femtocell coverage that corresponds to the building area, i.e.,
$r_f=r_b$. It is important to note that this method is
effective irrespective of the building's size, because $\Gamma_0=\frac{5n}{\ln10}+\gamma_{th}$ and
$\Gamma^{(i)}$ given by (\ref{eq:Gamma2}) or
(\ref{eq:Gamma3}) are independent of $r_b$.

\subsubsection{Maximum Additional Threshold $\Gamma_{\Delta,\max}$}
From Fig. \ref{fig:Gamma}, we can observe that when the threshold
$\Gamma_{th}$ higher than $\Gamma_0$ does not increase transmit power $P_f^{(i)}$ up to the level at which femtocell coverage leaks into an outdoor area, a downlink
CINR of femtocell is improved while $r_f^{(i)}=r_b$. From this observation we define $\Gamma_{\Delta,\max}$ as the maximum increase, which satisfies the condition $r_f^{(i)}=r_b$, from the basic threshold $\Gamma_{0}$.

$\Gamma_{\Delta,\max}$ is
designed from the two cases\footnote{In this section, several parameters ($\Gamma^{(i)}$,
$Q^{(i)}$, $r_f^{(i)}$, and $P_f^{(i)}$) defined in the previous
section are classified into the parameters of Cases 1 and 2 using
subscript numbers $_1$ and $_2$, and the iteration index $^{(i)}$ is
abbreviated for notational convenience} of femtocell coverage described in Fig. \ref{fig:Margin}. In Case 1, femtocell coverage is extended to a
building wall, i.e., $r_f=r_b$, by using $\Gamma_{th}=\Gamma_0$.
Increasing the transmit power of femtocell BS more than that of Case
1 results in Case 2, in which the femtocell coverage begin to leak into
an outdoor region, i.e., $r_f=r_b+\Delta D$, where $\Delta D$ is a
very small. Thus $\Gamma_{\Delta,\max}$ is given as
\begin{equation}\label{eq:GammaDeltaMax}
\Gamma_{\Delta,\max}=\Gamma_{2}-\Gamma_{1}
\end{equation}
Note that while the femtocell BS
increases the transmit power from Case 1 to Case 2, $r_f$ remains
equal to $r_b$, and $\overline{Z}(r_f)$ is invariant. From (\ref{eq:Gamma1}), we then obtain
\begin{eqnarray}\label{eq:GammaDeltaMax1}
\Gamma_{2}-\Gamma_{1}&=&E[Q_2]-E[Q_1]\cr
&\mathop =\limits^{\left( a \right)}&P_{f,2}-P_{f,1}\cr
&\mathop =\limits^{\left( b \right)}&Q_{2}(r_b)-Q_{1}(r_b)
\end{eqnarray}
where (a) follows from (\ref{eq:E_R}) and (b) follows from the equation
$Q(D)=P_f-L\left(D\right)$.
In Case 1, as $\Gamma_{1}=\Gamma_0$ and the CINR constraint
$Q(r_b)-\overline{Z}(r_b)=\gamma_{th}$ is preserved, the received
pilot power of the femtocell at $D=r_b$ is given by
\begin{equation}
Q_{1}(r_b)=\gamma_{th}+\overline{Z}(r_b)\label{eq:Qin}
\end{equation}
In Case 2, the boundary between the femtocell and macrocell is
defined as the position where the received pilot powers of both
cells are identical. Therefore, $Q_{2}(r_b+\Delta
D)=I_{\max}(r_b+\Delta D)$ at $D=r_b+\Delta D$, where $I_{\max}$ is
the received pilot power from the strongest interfering BS. When
this condition is satisfied, $Q_{2}(r_b)$ is given as
\begin{equation}
Q_{2}(r_b)=I_{\max}(r_b)+2L_p\label{eq:Qout}
\end{equation}
Combining (\ref{eq:GammaDeltaMax}), (\ref{eq:GammaDeltaMax1}), (\ref{eq:Qin}), and (\ref{eq:Qout}),
$\Gamma_{\Delta,\max}$ is given as
\begin{equation}
\Gamma_{\Delta,\max}=I_{\max}(r_b)+2L_p-\gamma_{th}-\overline{Z}(r_b)
\label{eq:mu_th}
\end{equation}
In conclusion, $\Gamma_{th}=\Gamma_0+\Gamma_{\Delta},~\textrm{where
} 0\leq \Gamma_{\Delta} \leq\Gamma_{\Delta,\max}$, provides a higher
downlink CINR than $\Gamma_{th}=\Gamma_0$ due to the additional
femtocell transmit power, while preserving the femtocell coverage
that corresponds to the area of the building.


\section{Femtocell Coverage Analysis}
\label{sec:coverage} The statistical threshold $\Gamma_0$ is derived
from the expected value $E\left[Q\right]$, but $\Gamma$, used for
the power control, is estimated by using the sample mean
$\overline{Q}$ at a femtocell BS. This results in a coverage
leakage. Therefore, the coverage leakage probability that femtocell
coverage leaks into the outdoor macrocell, $H_K$, is derived in this
section. It is assumed that $K$ femtocell users are uniformly
distributed in a building, and one of them is located at the
boundary of the building, $D=r_b$.
The received pilot power averaged over $K$ femtocell users is given
by
\begin{equation}
\overline{Q}=\left(\textstyle{\sum_{k=1}^{K-1}}Q_k+Q_K
\right)\Big/K=\left(\textstyle{\sum_{k=1}^{K-1}}Q_k+(P_f-A_s-10n\log(r_b))
\right)\Big/K,\label{eq:Def_M_N}
\end{equation}
where $Q_k$ and $Q_K$ is the received pilot power of the $k$th
femtocell user and the femtocell user located at $D=r_b$,
respectively. When $r_f$ has approached $r_b$, the $\Gamma$
estimated in a femtocell BS is given as
$\Gamma=\overline{Q}-\overline{Z}(r_b)$, and the femtocell BS
increases its transmit power until $\Gamma$ increases to
$\Gamma_{th}$. If the additional femtocell transmit power, which is
estimated to be $\Gamma_{th}-\Gamma$, is greater than
$\Gamma_{\Delta,\max}$, the femtocell coverage leaks into an outdoor
area. Thus, $H_K$ is defined and determined by using
$\Gamma_{th}=\Gamma_{0}+\Gamma_{\Delta}$, $\Gamma_{0}=
E[Q]-\overline{Z}(r_b)=E[Q]-Q_K+\gamma_{th}$, and
$Z_u(0)\approx\overline{Z}(r_b)$, as follows:
\begin{eqnarray}
H_K&\triangleq&\mathrm{Pr}\left[\Gamma_{th}-\Gamma >
\Gamma_{\Delta,\max}\right]\cr
&\approx&
\mathrm{Pr}\left[\overline{Q}<E[Q]+\Gamma_{\Delta}-\Gamma_{\Delta,\max}\right]\cr
&=& \mathrm{Pr}[\overline{Q}-Q_K<
\Gamma_0-\gamma_{th}+\Gamma_{\Delta}-\Gamma_{\Delta,\max}].\label{eq:Def_H}
\end{eqnarray}

Let the random variable $X$ be defined as $X=\left(Q-Q_K\right)/K$.
Then, from the PDF of $Q$, the PDF of $X$ is given as
\begin{equation}
f_X(x)=\lambda_2e^{-\lambda_1x},~x\in\left[0,10n\log\left(r_b/\varepsilon_0\right)/K\right],
\label{eq:PDF_X}
\end{equation}
where $\lambda_1=K\ln10/5n$,
$\lambda_2=K\ln10\cdot10^{\frac{Q_K+10n\log(r_b)}{5n}}e^{-\frac{\lambda_1}{K}Q_K}/(5n\left(r_b^2-\varepsilon_0^2\right))=r_b^2K\ln10/(5n(r_b^2-\varepsilon_0^2))$.
As $\lambda_2\approx\lambda_1$ for $r_b^2\gg\varepsilon_0^2$, the
PDF of $X$ is approximated to that of the exponential random
variable with a parameter of $\lambda_1$, i.e.,
$f_X(x)\approx\lambda_1e^{-\lambda_1x}$.
Next, let denote $Y_M$ as the sum of $M$ independent, identically
distributed random variables $\{X_m\}_{m=1,\cdots,M}$ with a PDF
identical to that of $X$:
\begin{equation}
Y_M=\textstyle{\sum_{m=1}^{M}}X_m=\textstyle{\sum_{m=1}^{M}}\left(Q_m-Q_K\right)/K\label{eq:Def_YM}
\end{equation}
The cumulative distribution function of $Y_M$ is then approximated
to that of an Erlang random variable that was obtained by adding $M$
independent exponential random variables with a parameter of
$\lambda_1$ as \cite{Garcia}
\begin{equation}
F_{Y_M}(y)\approx
1-\sum_{m=0}^{M-1}\frac{e^{-\lambda_1y}(\lambda_1y)^{m}}{m!}
,~y\in\left[0,10nM\log\left(r_b/\varepsilon_0\right)/K\right].
\label{eq:CDF_YM}
\end{equation}
From (\ref{eq:Def_M_N}) and (\ref{eq:Def_YM}),
$Y_{K-1}=\overline{Q}_K-Q_K$, and
$y_0=\Gamma_0-\gamma_{th}+\Gamma_{\Delta}-\Gamma_{\Delta,\max}$;
(\ref{eq:Def_H}) is rewritten as
\begin{eqnarray}
H_K&=&\mathrm{Pr}\left[Y_{K-1}<y_0\right]\cr
&=& F_{Y_{K-1}}(y_0)\cr
&\approx&
1-\textstyle{\sum_{m=0}^{K-2}}e^{-\lambda_1y_{0}}(\lambda_1y_0)^m\big/m!~,~
\textrm{for}~~y_0\geq 0\label{eq:H_K}
\end{eqnarray}
and $H_K=0$ for $y_0<0$.

In (\ref{eq:H_K}), $e^{-\lambda_1y_{0}}(\lambda_1y_0)^m/m! \geq0$
because $y_0\geq0$. Thus, $H_K$ decreases as $K$ increases, which
demonstrates that a larger number of femtocell users improve the
performance of the proposed scheme. Additionally, from the Taylor
series for the exponential function $e^x=\sum_{m=0}^{\infty}x^m/m!$,
the asymptotic behavior of $H_K$ at a very large $K$ is given as
\begin{eqnarray}
\lim_{K\rightarrow\infty}
H_K=1-e^{-\lambda_1y_{0}}\sum_{m=0}^{\infty}\frac{(\lambda_1y_0)^m}{m!}=
0.\label{eq:Def_Hinf}
\end{eqnarray}
This indicates that the proposed scheme is asymptotically optimal in
terms of coverage leakage probability.



\section{Performance Evaluation}\label{sec:results}
\subsection{Simulation assumptions and performance metrics}
With the system parameters given in Table I, a Monte Carlo
simulation approach is used to evaluate the coverage leakage
probability $H_K$, the average leakage distance $\Omega$, and the
average indoor coverage $\Psi$. A macrocell has a layout of 19
hexagonal cells arranged in a hexagonal lattice with two rings of
cells surrounding the center cell. A target femtocell is located at
a point with distance $d_b=400$ m from the macrocell BS in the
center cell, and 50 interfering femtocells with a fixed pilot power
given from (\ref{eq:InitPw}) are uniformly distributed within the
center macrocell with a radius of $r_m=580$ m. For each simulation
repetition in a Monte Carlo simulation with 5,000 trials, the indoor
and outdoor users are uniformly distributed in the building and the
outdoor area from 10 m of the building wall, respectively. After
the geometrical configuration of the BS and users, the femtocell
pilot power is initiated from (\ref{eq:InitPw}) and is optimized by
using (\ref{eq:Update1}) until it converges. The femtocell coverage
is then evaluated so that the received pilot power of femtocell is
larger than that of the macrocell. $H_K$ is obtained by dividing the
number of events where femtocell coverage leaks outdoors by the
total number of trials. $\Omega$ is evaluated by averaging the
distance between the leaked femtocell edge and building wall over
the total number of trials. $\Psi$ is calculated as the ratio of
average femtocell coverage to the building's area. $H_K$ and
$\Omega$ measure femtocell coverage that leaks into the outdoor
area. Additionally, $\Psi$ measures the performance of providing
sufficient indoor femtocell coverage.

\subsection{Simulation results}
Fig. \ref{fig:Result1} shows $H_K$ versus $\Gamma_{\Delta}$ for
$K$=5, 10, 20, 40, and infinity when $L_p$ = 10 and 3 dB. The
analytic curves are very close to the simulated curves. $H_K$
increases as $\Gamma_{\Delta}$ increases due to the fact that the
higher $\Gamma_{th}$ that is induced by an increase in
$\Gamma_{\Delta}$ causes a rise in the transmit power of a femtocell
BS. A larger wall penetration loss $L_p$ reduces the $H_K$ due to
the increasing value of $\Gamma_{\Delta,\max}$ as shown in
(\ref{eq:mu_th}). Moreover, $\Gamma_{\Delta}$ guaranteeing some
level of $H_K$ increases as $L_p$ becomes larger, which results in
the higher CINR of a femtocell downlink. Therefore, the proposed
scheme is more effective for a building with a higher wall
penetration loss. Fig. \ref{fig:Result1} also shows the impact of
$K$ on $H_K$. A larger $K$ increases the probability that the sample
mean $\overline{Q}$ becomes close to the expected value of $E[Q]$,
i.e., the approximation error of (\ref{eq:Gamma2}) decreases, which
results in less $H_K$ where $H_K$ is zero for infinite $K$. A
description of this result is also in the last paragraph of Section
\ref{sec:results}. At least 40 users (less users are probably
deployed in a femtocell except for the enterprise scenario), should
be deployed in order to achieve an $H_K$ value of 5~\% when $L_p$ =
3 dB. However, according to available literature
\cite{WallLoss1}\cite{WallLoss3}, the probability that $L_p=3$ is
low in actual applications. Thus, the proposed scheme remains
preferable. The analysis and simulation of $H_K$ as shown in Fig.
\ref{fig:Result1} are performed by considering no shadowing and the
perfect estimation of path loss exponent $n$. Although this
assumption is unrealistic, it provides good insight regarding the
proposed algorithm's performance according to many factors of
$\Gamma_{th}$, $L_p$, $n$, and $K$.

In a real femtocell scenario we consider shadowing, path loss exponent
estimation error, a small number of users, and rectangular-shaped building with several non-symmetric wall penetration loss.
Although shadowing is included in the path loss model, $\Gamma_0$
does not change because the shadowing averaged over an indoor area
is zero, i.e., $E[Q]$ does not vary. On the other hand, more uneven
cell coverage, due to both shadowing and several non-symmetric wall penetration loss, highly increases $H_K$, but despite
of this higher $H_K$, the leakage area can be small so that few
outdoor users are linked to a femtocell BS. Thus, average leakage distance $\Omega$ and average indoor coverage $\Psi$ are investigated for the real scenario with parameters in Table I, and the results are shown in Fig. \ref{fig:Result2}. The path loss exponent error is considered by
using $n_{e}$ and $n_{r}$, which denotes the path loss exponent $n$ used for determining the
$\Gamma_0$ and calculating the received power in a real link,
respectively. A higher $\Gamma_{\Delta}$ leads to a rise in the
transmit power of the femtocell BS, which increases $\Omega$ and
$\Psi$. Thus, $\Gamma_{\Delta}$ is adaptively determined according
to both the maximum permissible $\Omega$ and minimum achievable
$\Psi$. From $\Gamma_0=\frac{5n}{\ln10}+\gamma_{th}$,
$n_{e}>n_{r}$ increase $\Gamma_{th}$, which gives the same impact
on the performance as $\Gamma_{\Delta}$ increases under the
condition $n_{e}=n_{r}$. On the contrary, $n_{e}<n_{r}$ leads to
opposite results. Thus, a higher, or a lower, $n_{e}$ is
recommended to increase $\Psi$ or decrease $\Omega$, respectively.
The simulation results indicate that the proposed algorithm archives $\Omega$ less than 5 m and $\Psi$ more than 0.9, which is a feasible level of
performance for the realistic scenario. Additionally, this scheme
requires additional uplink overhead for reporting the average
received pilot power. As the power is averaged over multiple frames
to remove fast fading, long-term reporting sufficiently supports the
amount of feedback information, i.e., the overhead is not so
considerable as to make implementation impossible.

\section{Conclusion}\label{sec:conclusion}
This paper proposes a novel coverage coordination scheme based on a
self-configuration and self-optimization of transmit power. An
analytic expression for the coverage leakage probability of the
femtocell is derived and verified by simulations. The simulation
results show that, by using the proposed scheme, femtocells provide
sufficient indoor coverage and low coverage leakage to outdoor area.
In conclusion, the proposed scheme can make femtocell coverage
correspond to the building's area without knowing about the area of
the building. Further research needs to improve the robustness
against the small number of users as well as to investigate the effect of mobility of users.

\begin{figure}
\begin{center}
   \includegraphics[width=4.in]{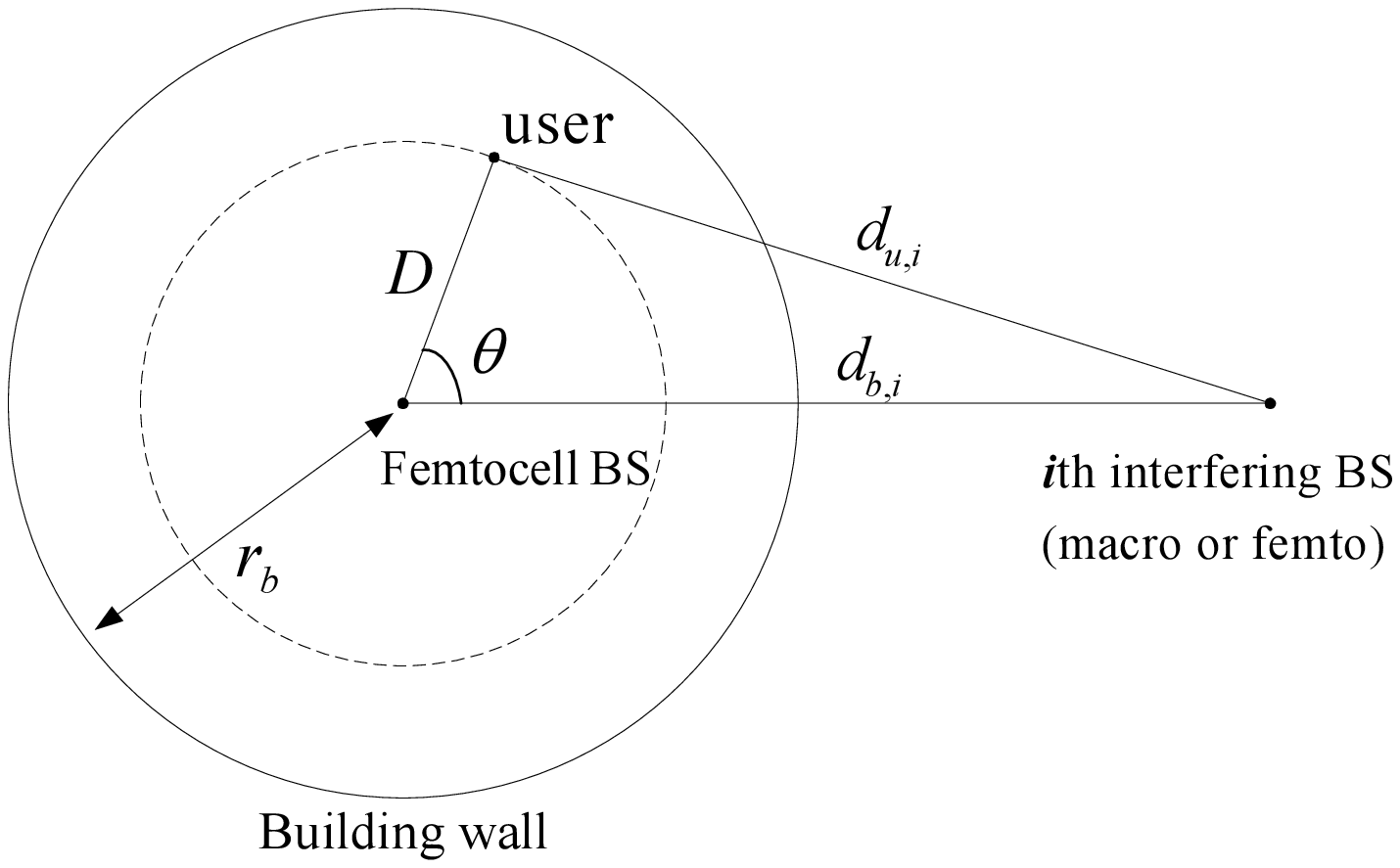}
    \caption{Geometric configuration for calculating the othercell interference at femtocell BS and users}
    \label{fig:Interference}
\end{center}
\end{figure}
\begin{figure}
\begin{center}
   \includegraphics[width=4.in]{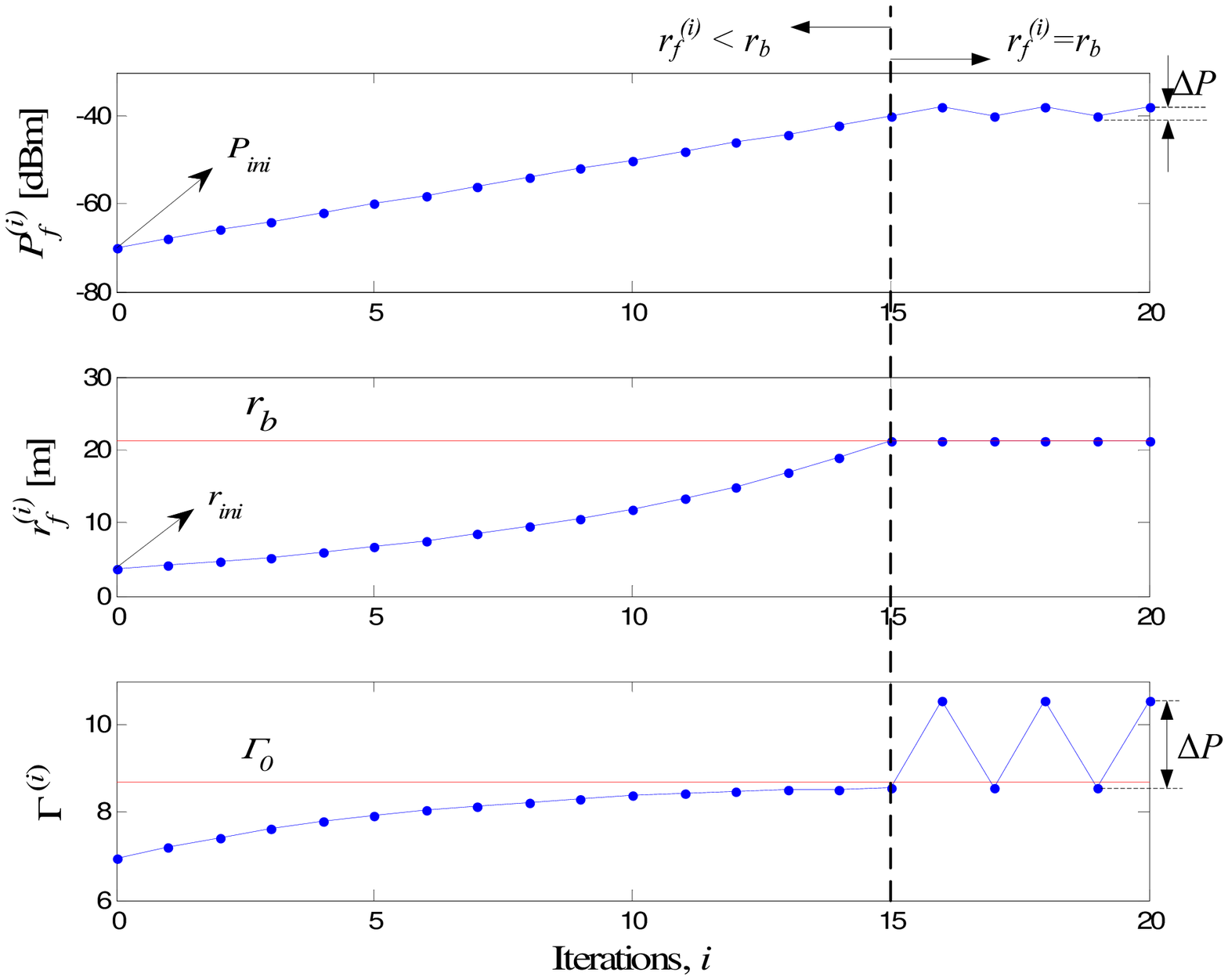}
    \caption{A change of transmit power, femtocell coverage, and $\Gamma^{(i)}$, using the proposed power control with $\Gamma_{th}=\Gamma_0$ and $\Delta P$=2 dB.}
    \label{fig:Gamma}
\end{center}
\end{figure}
\begin{figure}
\begin{center}
   \includegraphics[width=4.in]{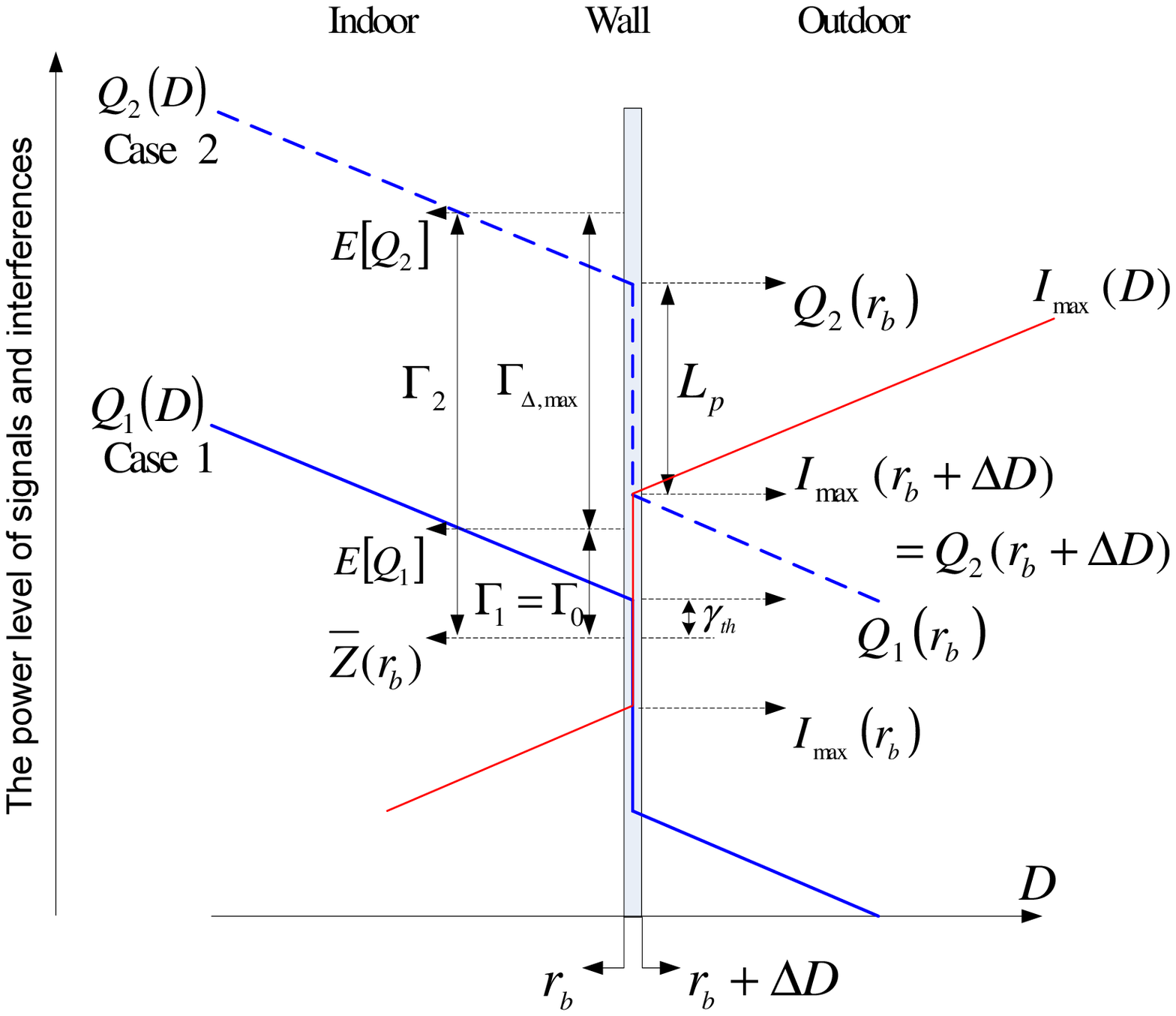}
    \caption{Wall penetration loss $L_p$ enables a femtocell BS to utilize the additional threshold
    $\Gamma_{\Delta}$, $0\leq\Gamma_{\Delta}\leq\Gamma_{\Delta,\max}$ (Case 1: $r_f=r_b$, Case 2: $r_f=r_b+\Delta D$).}
    \label{fig:Margin}
\end{center}
\end{figure}
\begin{figure}[h]
\begin{center}
   \includegraphics[width=4.in]{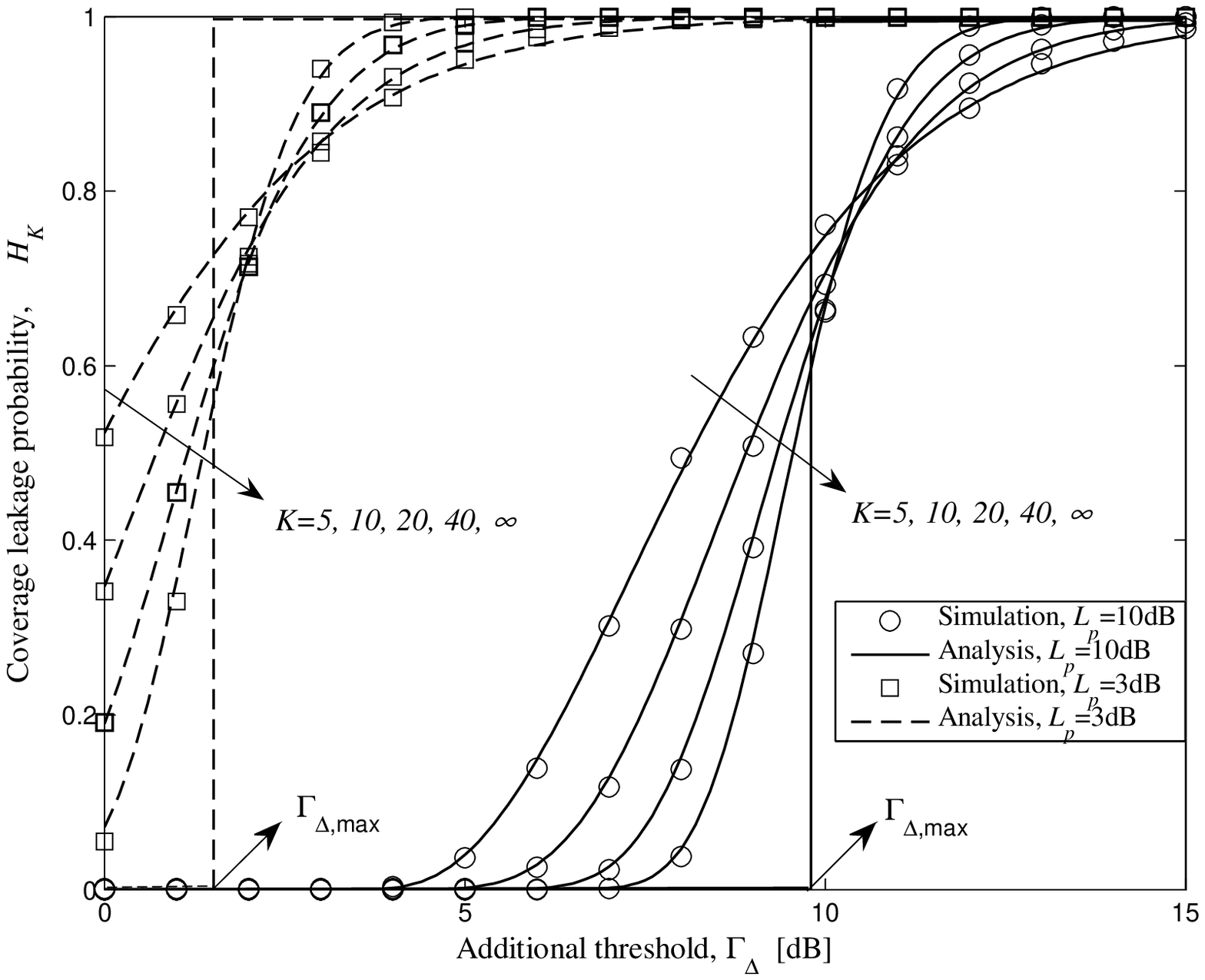}
    \caption{$H_K$ versus $\Gamma_\Delta$, comparing the simulation and analysis results for $L_p$ = 10 and 3 dB.}
    \label{fig:Result1}
\end{center}
\end{figure}
\begin{figure}[h]
\begin{center}
   \includegraphics[width=4.5in]{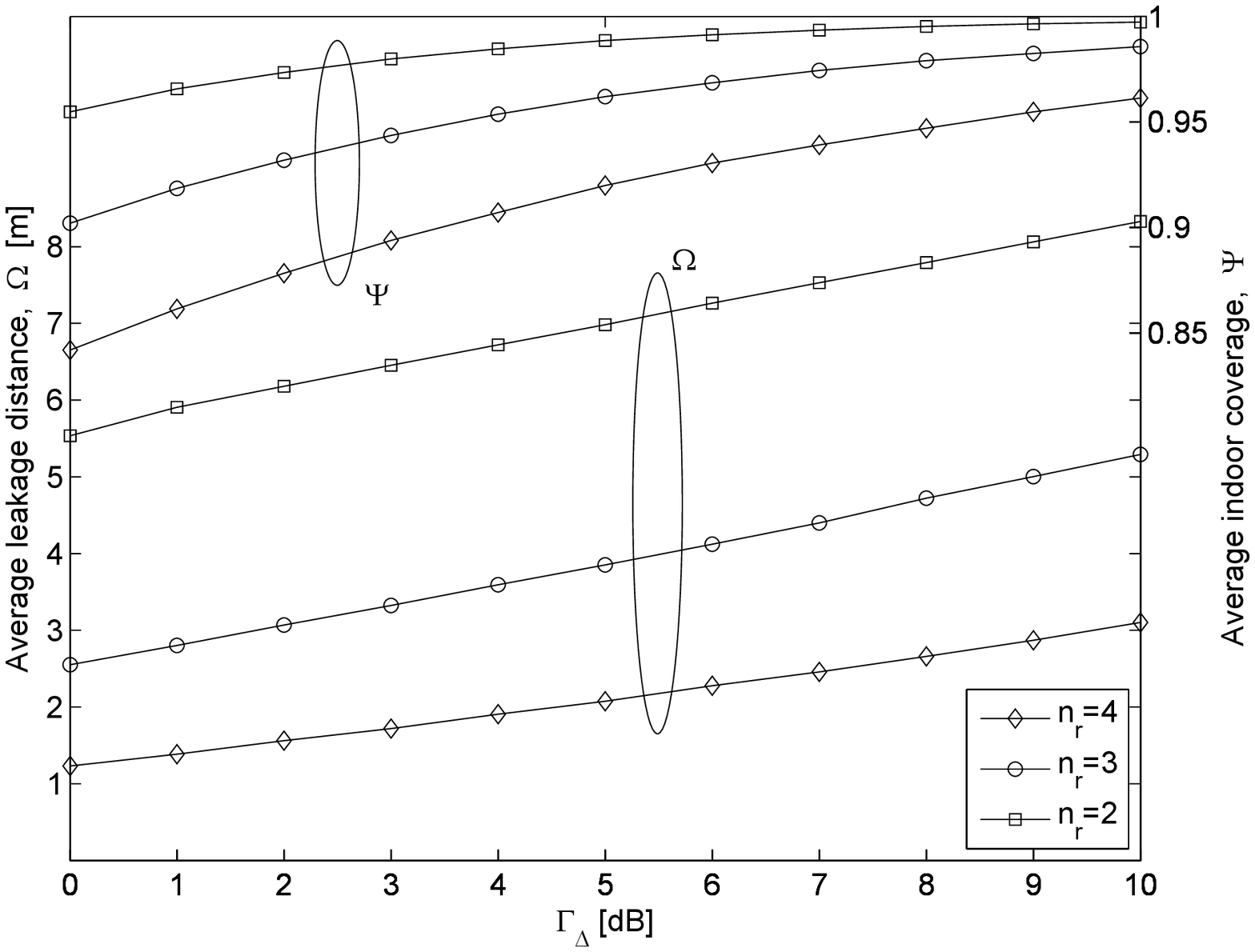}
    \caption{$\Omega$ and $\Psi$ versus $\Gamma_\Delta$ with shadowing , path loss exponent error ($n_e=3$), small number of users ($K=2$), and several non-symmetric wall penetration loss.}
    \label{fig:Result2}
\end{center}
\end{figure}

\appendix[]
Fig. \ref{fig:Interference} shows the geometric configuration for
calculating receiving power of the other-cell interference at a
femtocell BS and at its users. It is assumed that all $M$
interfering BSs use the same transmit power level $P_m$ in dBm and
that the wall penetration loss $L_p$ is constant irrespective of
$\theta$. The other-cell interference measured by and averaged over
the users, which are assumed to be located uniformly on a
circumference with a radius of $D$, is given by
\begin{eqnarray}\label{eq:I_u}
\overline{I}_u(D)&=&\frac{1}{2\pi}\int_{0}^{2\pi}
\sum_{i=1}^{M}\frac{P_{m,\mathrm{lin}}}{A_{s,\mathrm{lin}}L_{p,\mathrm{lin}}
(D^2+d_{b,i}^2-2Dd_{b,i}\cos\theta)^{-n/2}} d\theta \cr
&=&\frac{P_{m,\mathrm{lin}}}{A_{s,lin}L_{p,\mathrm{lin}}}
\sum_{i=1}^{M}(D+d_{b,i})^{-n}~
_2F_1\left[\frac{1}{2},\frac{n}{2};1;\frac{4Dd_{b,i}}{(D+d_{u,i})^2}\right],
\end{eqnarray}
where the subscript $\mathrm{lin}$ represents a linear value, and
$_2F_1[\cdot]$ is Gauss's hypergeometric function \cite{HyperGeo}.
$\overline{I}_u(D)$ can be obtain in a further simple form when
$n=2k$, as follows:
\begin{equation}\label{eq:I_uResult}
\overline{I}_u(D)=\frac{P_{m,\mathrm{lin}}}{A_{s,\mathrm{lin}}L_{p,\mathrm{lin}}}
\sum_{i=1}^{M}(D+d_{b,i})^{-3k+1}(D^2+d_{b,i}^2)^{k-1}
\Big/\left(1-\frac{4Dd_{b,i}}{(D+d_{b,i})^2}\right)^{\frac{2k-1}{2}},
\end{equation}
where $k$ is a integer \cite{HyperGeo}. On the other hand, the
other-cell interference at the femtocell BS is given by
\begin{equation}\label{eq:I_b}
I_u(0)=\textstyle{\frac{P_{m,\mathrm{lin}}}{A_{s,\mathrm{lin}}L_{p,\mathrm{lin}}}\sum_{i=1}^{M}d_{b,i}^{-n}}.
\end{equation}
Let $U$ and $B$ denote the summation part of (\ref{eq:I_uResult})
and (\ref{eq:I_b}), respectively. If $n=2,4$ and $d_{b,i}=2D,
3D,\textrm{and}~4D$ for all $i$, $U$ and $B$ are given as follows:
\begin{equation}\label{eq:UB2} U=\left\{\begin{array}{@{\hskip0pt} ccc @{\hskip0pt}}
\frac{0.333M}{D^2}, ~(\frac{0.185M}{D^4}) \cr \frac{0.125M}{D^2},
~(\frac{0.019M}{D^4}) \cr \frac{0.067M}{D^2}, ~(\frac{0.005M}{D^4})
\end{array}\right.,~~
B=\left\{\begin{array}{@{\hskip0pt} ccc @{\hskip0pt}}
\frac{0.25M}{D^2}, ~(\frac{0.063M}{D^4}), & d_{b,i}=2D \cr
\frac{0.111M}{D^2}, ~(\frac{0.012M}{D^4}), & d_{b,i}=3D \cr
\frac{0.063M}{D^2}, ~(\frac{0.004M}{D^4}), & d_{b,i}=4D
\end{array}\right. \textrm{for}~n=2,(\textrm{or }4)
\end{equation}
The similar values of $U$ and $B$ shows that 
$\overline{I}_u(D)\approx I_u(0)$ for the realistic conditions of a
path loss exponent $n$ larger than 2 and $d_{b,i}>2D$.



\newpage
\begin{table}
\caption{System parameters for the simulation and analysis}
\label{table_1}
\begin{center}
\begin{tabular}{c|c|c}
  \hline
  Parameter & Symbol & Value \\ \hline \hline
  Macrocell radius & $r_m$ & 580 m \\ \hline
  Macro-to-femto BS distance & $d_b$ & 400 m \\ \hline
  Building shape & & Theory verification (Fig. 4): Circle \\
  &&with a radius of $r_b=20\mathrm{m}$,\\
  & & Real scenario (Fig. 5): \\
  &&$20\mathrm{m} \times 15\mathrm{m}$ rectangle \\ \hline
  Wall penetration loss && Theory verification (Fig. 4):$L_p=3, 10 \mathrm{dB} $\\
  (Percent value represents &&Real scenario (Fig. 5): \\
  wall-length ratio) &&15dB(35\%), 10dB(30\%), 7dB(20\%), 2dB(15\%)\\ \hline
  Initial femtocell radius & $r_{ini}$ & 15 m \\ \hline
  Minimum distance & $\varepsilon_0$ & 1 m \\
  between a femtocell BS and a user  & \\ \hline
  Path loss at 1 m & $A_s$ & 37 dB \\ \hline
  Power control step & $\Delta P$ & 0.25 dB \\ \hline
  Maximum transmit power of a BS & $P_{\max}$, $P_m$ & 23 dBm (femto), 43 dBm (macro) \\ \hline
  Thermal noise power & $W$ & -96.8 dBm \\ \hline
  Path loss exponent & $n$ & 3 (femtocell), 4 (macrocell) \\ \hline
  CINR threshold & $\gamma_{th}$ & -2.6 dB \\ \hline
  Statistical threshold & $\Gamma_0$ & 3.91 \\ \hline
  Maximum additional threshold & $\Gamma_{\Delta,\max}$ & 1.52 dB ($L_p$=3), 9.8 dB ($L_p$=10)\\ \hline
\end{tabular}
\end{center}
\end{table}
\newpage

\end{document}